\documentstyle[preprint,aps,amstex,floats,tighten]{revtex}
\input epsf
\begin{document}
\pagestyle{myheadings}

\title{Production and decay of spinning black holes at colliders and \\ 
tests of black hole dynamics}
\author {Ashutosh V. Kotwal and Christopher Hays \\ {\em Department of 
 Physics, Duke University, 
Durham, NC 27708-0305, USA} } 
\maketitle
\vspace*{-0.1in}
\begin{abstract}
We analyse the angular momentum distribution
 of black holes produced in high energy collisions in space-times
 with extra spatial dimensions. We show that the black hole spin 
 significantly   affects the energy and angular
 spectra of Hawking radiation. Our results show the experimental
 sensitivity to the angular momentum distribution and provide tests
  of black hole production dynamics. 
\end{abstract}




 The possibility of extra spatial dimensions (ESDs) has been 
 raised to explain the heirarchy problem, either through the volume
 of the ESDs~\cite{arkanihamed} or an exponential warp factor~\cite{rs}. 
 Given a number $n$ of ESDs,  
 the hypothesis is that gravity propagates in $4+n$ dimensions (the
 bulk), while  the standard model (SM) fields are confined to the
 4 observed dimensions (the brane). 
 At the fundamental scale $M_{F}$ in $4+n$ dimensions, which
 is close to the electroweak scale $\cal O$(1 TeV), all interactions
 including gravity have the same strength.  

 In a parton-parton collision with energy exceeding $M_{F}$,
  the strong gravitational field can induce the formation of a
 black hole if the energy is confined to a region smaller than the
 corresponding event horizon~\cite{banksfischler}. 
  The production and decay of black holes of mass $ M_{bh} \sim M_{F}$
 will be heavily influenced by quantum gravitational effects.
  However, for  $M_{bh} \gg M_{F}$, the classical picture 
 based on Einstein's equations of general relativity 
should be valid. The analysis of the behavior of
 quantum fields in the classical metric
 results in the Hawking effect~\cite{Hawking}: the black hole radiates
 energy with a spectrum characteristic of a black body.  
 The resulting 
 thermodynamic description~\cite{Hawking2}
  identifies the black hole surface gravity and the surface area
 with the temperature and the entropy, respectively, of the black body.

The Hawking radiation from black holes provides information about their 
properties~\cite{giddingsthomas,dl}. 
  In this Letter 
 we show that the angular momentum is an important parameter 
 in the production and decay of collider-produced black holes. 
 In a purely geometrical picture,
 partial wave analysis indicates that black holes are
 preferentially produced with large angular momentum. 
 We study the properties 
 of Hawking radiation from spinning black holes, and
 propose two measurements that test 
 various production models. These measurements are relevant
 for understanding the role played by the black hole entropy
 and the Gibbons-Hawking action in the production dynamics. 

In $4+n$ dimensions, the radius of the black hole event horizon 
 $R_{bh}$ is given
 by \cite{myersperry}
\begin{equation}
R_{bh}^{n+1} ( 1 + a_*^2 ) = \frac {16 \pi M_{bh}} {(n+2)A_{n+2} M_{F}^{n+2} } \quad ,
\label{radeqn}
\end{equation}
where $ A_{n+2} = 2 \pi ^ {(3+n)/2} / \Gamma ( \frac {3+n} {2} )  $
 is the area of the unit
 $n+2$ sphere,  and
$ a_* =  (n+2) J / ( 2 M_{bh} R_{bh} ) $
  is a dimensionless rotation
 parameter for the 
 angular momentum $J$ of the black hole. We use the
 convention $\hbar = c = k_B = 1$, where $ k_B$ is the Boltzmann constant, and
 the gravitational constant $ G = M_{F}^{-(n+2)}  $. 

The size $L$ of the ESDs is related to $M_{F}$
 by  $L \sim (M_{Pl}^2/M_{F}^{n+2})^{1/n}$ \cite{arkanihamed}, where 
$M_{Pl}$ ( $\sim 10^{16}$ TeV ) is the 4-dimensional Planck scale.
   The horizon radius $R_{bh}\sim (M_{bh}/M_{F}^{n+2})^{1/(n+1)}$ 
 is much smaller than $L$ when
\begin{equation}
 M_{bh}^{\frac{n}{n+1}}M_{F}^{\frac{n+2}{n+1}} \ll M_{Pl}^2 \quad .
\end{equation}
This is the case for the collider-produced black holes we consider, so
these black holes ``live'' in $4+n$ dimensions. 

 In the center-of-mass (CM) frame of a two-parton
 collision,
 the energy of each parton is $M_{bh}/2$. The geometrical transverse phase 
 space is maximized for parton impact parameter near $R_{bh}$, implying that
 $J  \sim R_{bh} M_{bh} \sim  (M_{bh}/M_{F})^{\frac {n+2} {n+1} } $. 
 In 4 dimensions ($n=0$), the initial state of the black hole tends to be 
 extremal, {\it i.e.} $J$ approaches the upper limit
 $M_{bh}^2 / M_{F}^2$.  
 
 The temperature of a spinning black hole is given by~\cite{myersperry} 
 \begin{equation}
T_{bh} =  \frac { n+1 + (n-1) a_*^2 } { 4 \pi R_{bh} (1+a_*^2) } \quad .
\label{temperature}
\end{equation}
 The energy ($\omega$) spectrum 
 of the radiated particles from a spinless black hole
 is given by the Planck spectrum of thermal black body 
 radiation~\cite{Hawking2}, for which the mean energy $\omega_*$ is 
   $\cal O$$ ( T_{bh}) $. The decay angular 
 distribution is affected by structure of size $r$, where 
 $ r \sim \omega_* ^{-1} \sim T_{bh} ^{-1}  \sim R_{bh} $.  
 Thus, for a typical spinning black hole produced in a high 
energy collision, we 
 expect the angular distribution of the Hawking radiation to deviate from an
 isotropic distribution. We discuss below the predicted black hole
 $J$ distribution, and the radiation $\omega$ and
 angular distributions,  based on a detailed Monte Carlo model of black hole
 production and decay. 

{\it Black hole production}: We compute the probability
  of producing a black hole with given $M_{bh}$ and $J$ by 
using the partial wave expansion of the initial state, 
 the interaction Hamiltonian and the 
black hole phase space. 
 We make two assumptions: (i) the Hamiltonian conserves angular momentum, 
and (ii) the Hamiltonian is not spin-dependent. 
We take the Hamiltonian to be non-zero when 
$|\; \vec{r}_1-\vec{r}_2|\leq 2 R_{bh}$, \em i.e. \em  
 $\hat{H} \sim \Theta(2 R_{bh}-| \; \vec{r}_1-\vec{r}_2 |)$, where 
 $\vec{r}_1$ and  $ \vec{r}_2 $ are the position vectors of the 
 colliding partons. 

We approximate the wave functions of the colliding partons 
by plane waves. We define the CM and relative position vectors to
be $\vec{R}=(\vec{r}_1+\vec{r}_2)/2$ and $\vec{r}=\vec{r}_1-\vec{r}_2$  
respectively, and we take $\vec{k}_{1,2} = \pm k\hat{z}$ to be the momenta of the 
incoming partons ($k = \omega$). 
Using the CM frame, we 
denote the two-parton initial state by $|~\Psi,~\psi,~\chi >$, 
where $<\vec{R}~|~\Psi> = A e^{i (\vec{k}_1 + \vec{k}_2) \cdot \vec{R}} = A$, 
  $<\vec{r}~|~\psi> = A' e^{i (\vec{k}_1 - \vec{k}_2) \cdot \vec{r}/2}$ and $\chi$ is the 
two-parton intrinsic 
 spin state. $A$ and  $A'$ are normalization factors that include a cut-off for the 
volume containing the plane waves.  
 We denote the black hole state by 
$<J~J_z~|$, where $J_z$ is the $z$ component of $J$.  The
probability of producing a black hole with angular momentum $J$ is
$|< J \; J_z~| \; \hat{H} \; | \; \Psi \; \psi \; \chi >|^2 \times \cal{F},$ 
where $\cal{F}$ is the black hole phase space. 
Using the plane wave expansion in spherical harmonics~\cite{planewave},  
 the matrix element 
$\cal{M}$=$< J \; J_z \; | \; \hat{H} \; | \; \Psi \; \psi \; \chi >$ reduces
 to
\begin{eqnarray}
{\cal M} &= & \int d^3 \vec{R} \; d^3 \vec{r} < J \; J_z \; | \; \hat{H} \; | \; \vec{R} 
\; \vec{r} \; \chi > <\vec{R} \; \vec{r} \; | \; \Psi \; \psi> \nonumber \\
 & = & A'' \sum_{lsm_s} ~ [~  <J~J_z~|~l~0~s~m_s>~\sqrt{2l+1} \nonumber \\
 & & \times~\int _{2R_{min}}^{2R_{bh}} r^2 dr ~j^2_l(\frac{M_{bh}}{2} r) ~] \quad ,
\end{eqnarray}
where $<J~J_z~|~l~0~s~m_s>$ are the Clebsch-Gordon coefficients and $s$ ($m_s$)
 is the total ($z$ component of the) two-parton intrinsic spin.  We assume
 that space-time is quantized in the underlying theory of quantum gravity,  
 setting the minimum black hole radius ($R_{min}$) 
 to $M_F^{-1}$.

 Another method~\cite{voloshin} of calculating the matrix element
 using the path-integral approach
 has introduced the Gibbons-Hawking action~\cite{gibbonshawking}.
 The results of this approach have been 
 debated~\cite{noentropy,solodukhin}.  
 We comment on this model at the end of our paper. 
 
\begin{figure}[hptb]
\begin{center}
\epsfxsize 9.0cm
\epsfysize 9.0cm
\epsffile{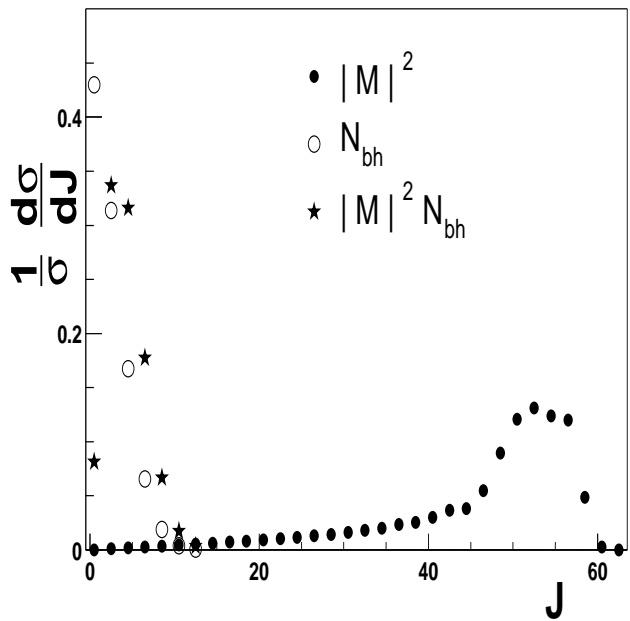}
\vskip 0.15in
\caption{The normalized 
 distributions of black hole angular momentum $J$, for 
 $M_{bh}=50 M_F$ in 4+3 dimensions.  We show the separate contributions of  
$|{\cal M}|^2$ and $N_{bh}$, and their product. }
\vskip -0.15in
\label{fig:angularmomentum}
\end{center}
\end{figure}

 One may take the black hole phase space $\cal{F}$ as
 the number of available black hole 
 states $N_{bh}$ \cite{fischler}, which is related to 
 the entropy $S_{bh} = \ln (N_{bh})$, given by~\cite{myersperry}
\begin{equation}
S_{bh}=\frac{4\pi R_{bh}M_{bh}}{n+2} \quad .
\end{equation}
However, this choice of ${\cal F}$ 
 contradicts the purely geometrical picture, where ${\cal F} = 1$.
 These two choices of ${\cal F}$ result 
 in dramatically different $J$ distributions.
  Figure~\ref{fig:angularmomentum} shows the normalized distributions
of $|{\cal M}|^2$, $N_{bh}$, and $ |{\cal M}|^2 N_{bh} $ as a function of $J$,
for a black hole of mass $M_{bh}=50 M_F$ produced by two colliding fermions 
in 4+3 dimensions.  As we will show, the different $J$ distributions
 result in different radiation
 spectra. 

{\it Black hole decay}:  The decay by Hawking radiation
 is prescribed by thermodynamics~\cite{Hawking2}.  
 In our Monte Carlo simulation, 
 we model the black hole decay 
 as the sequential emission of individual partons~\cite{finaldecay}.  
 We derive 
 the $\omega$
  and angular momentum ($\vec{j}$) spectra
 using the micro-canonical ensemble~\cite{casadio} 
\begin{equation}
{\cal P}(\omega,\vec{j}) = \frac {g \; e^{\Delta S (\omega,\vec{j}) } }
{ \sum_i g_i \; e^{\Delta S (\omega_i,\vec{j}_i) } } \quad ,
\label{eq:blackbodyradiation}
\end{equation}
 where $g$ is the degeneracy of parton states, 
 $\Delta S$ is the change in entropy of the black hole, and the sum in
 the denominator 
 is over accessible parton $(\omega,\vec{j})$ values.
  The use of the micro-canonical ensemble
 incorporates finite mass effects such as the variation of the temperature
 and spin of the black hole during the decay. Since the black
hole entropy always increases when matter falls into the black hole 
\cite{Hawking2}, time reversal invariance requires $\Delta S$ to be negative 
in the black hole decay.  We omit in Eqn.~\ref{eq:blackbodyradiation} a 
 multiplicative grey-body factor $\Gamma _{\omega \vec{j}}$, which would
 represent the 
 absorption probability of a
 parton of given $\omega$ and $\vec{j}$.  A differential
 equation for $\Gamma _{\omega \vec{j}}$
 has been derived~\cite{teukolsky,greybody}, which suggests
  that high $j$ decays are suppressed for low energy radiation 
due to the grey body factor.  Including this general 
behavior of $\Gamma _{\omega \vec{j}}$
 in Eqn.~\ref{eq:blackbodyradiation} would further enhance
 the effects of black hole spin on the decay distributions that 
we discuss here.

\begin{figure}[hptb]
\begin{center}
\epsfxsize 9.0cm
\epsfysize 9.0cm
\epsffile{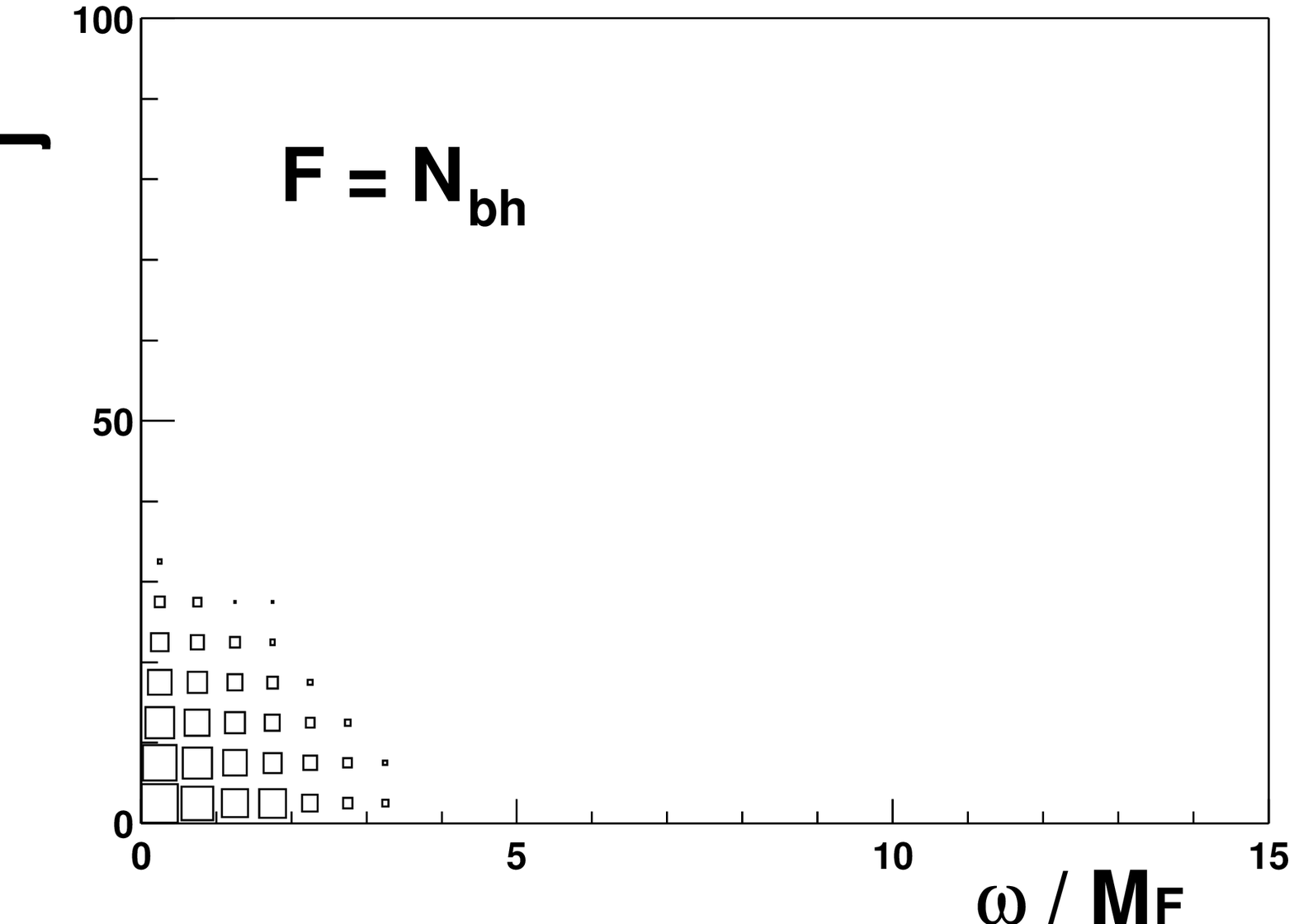}
\epsfxsize 9.0cm
\epsfysize 9.0cm
\epsffile{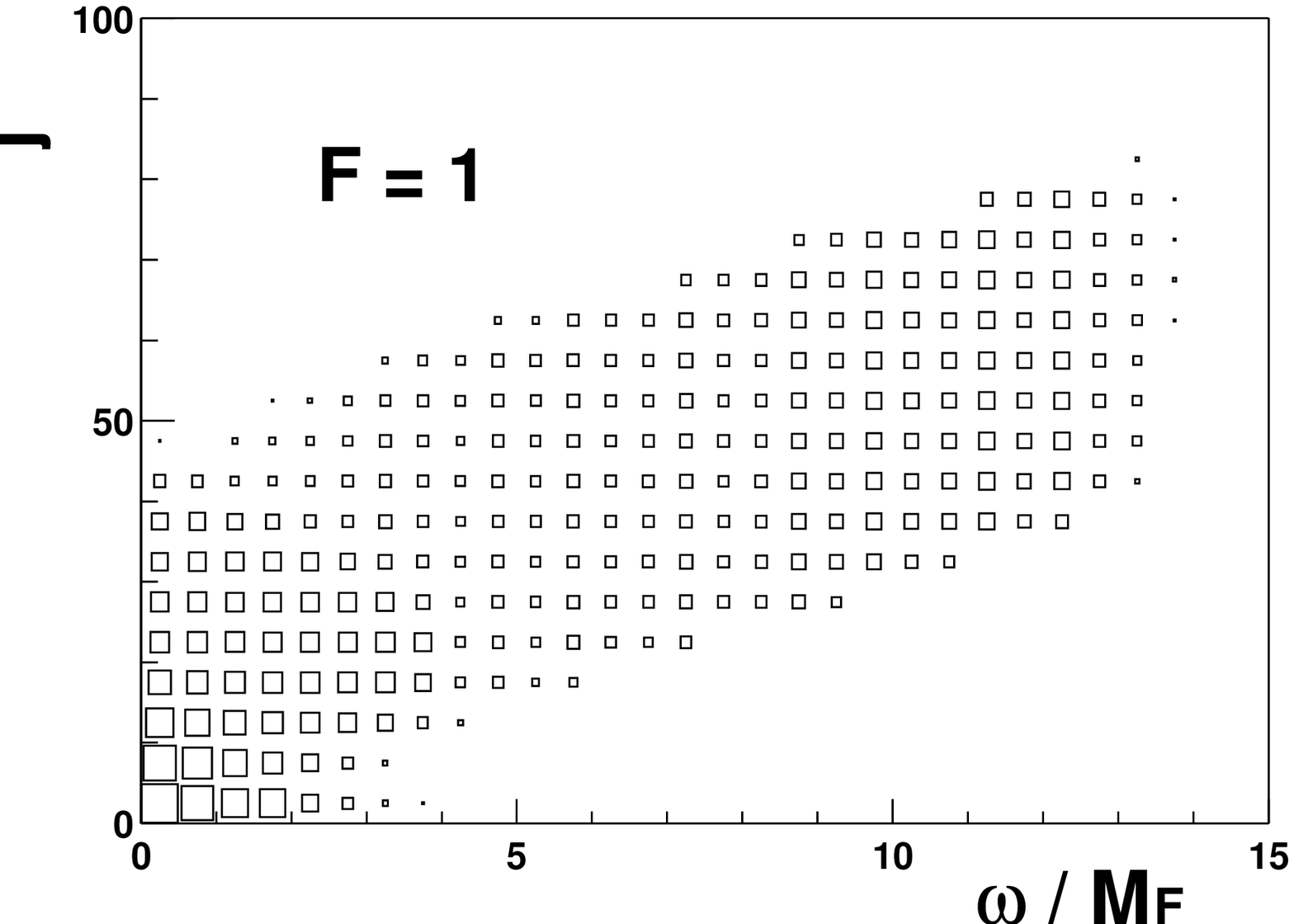}
\vskip 0.15in
\caption{The distribution of angular momentum versus energy of the
 decay partons for 
the two choices of $\cal{F}$ in the black hole production model.
  The distributions are shown for $n=3$ and $M_{bh} = 50 M_F$. 
 The density of points is indicated on a logarithmic scale. }
\vskip -0.15in
\label{fig:phasespacecomparelenergy}
\end{center}
\end{figure}

The exponential term $e^{\Delta S}$ in $\cal P$
 prevents large changes in black hole entropy in the
 emission of each parton.  Since 
the entropy decreases with decreasing mass, but increases with decreasing 
angular momentum, radiation of high energy partons
 is not suppressed if the partons also carry 
 high angular momentum.  
Therefore a spinning black hole emits more high energy partons
 compared to a spinless black hole of the same mass. 
 This property can 
 be used to distinguish between production models with different 
 choices of the phase space factor $\cal F$.
  The distributions of angular momentum versus energy of the
 radiated partons  are shown in Fig.~\ref{fig:phasespacecomparelenergy}. 
 When ${\cal F} = 1$, the black
hole radiates partons with higher $j$, and thus higher $\omega$,
 compared to ${\cal F} = N_{bh}$.

\begin{figure}[hptb]
\begin{center}
\epsfxsize 9.0cm
\epsfysize 9.0cm
\epsffile{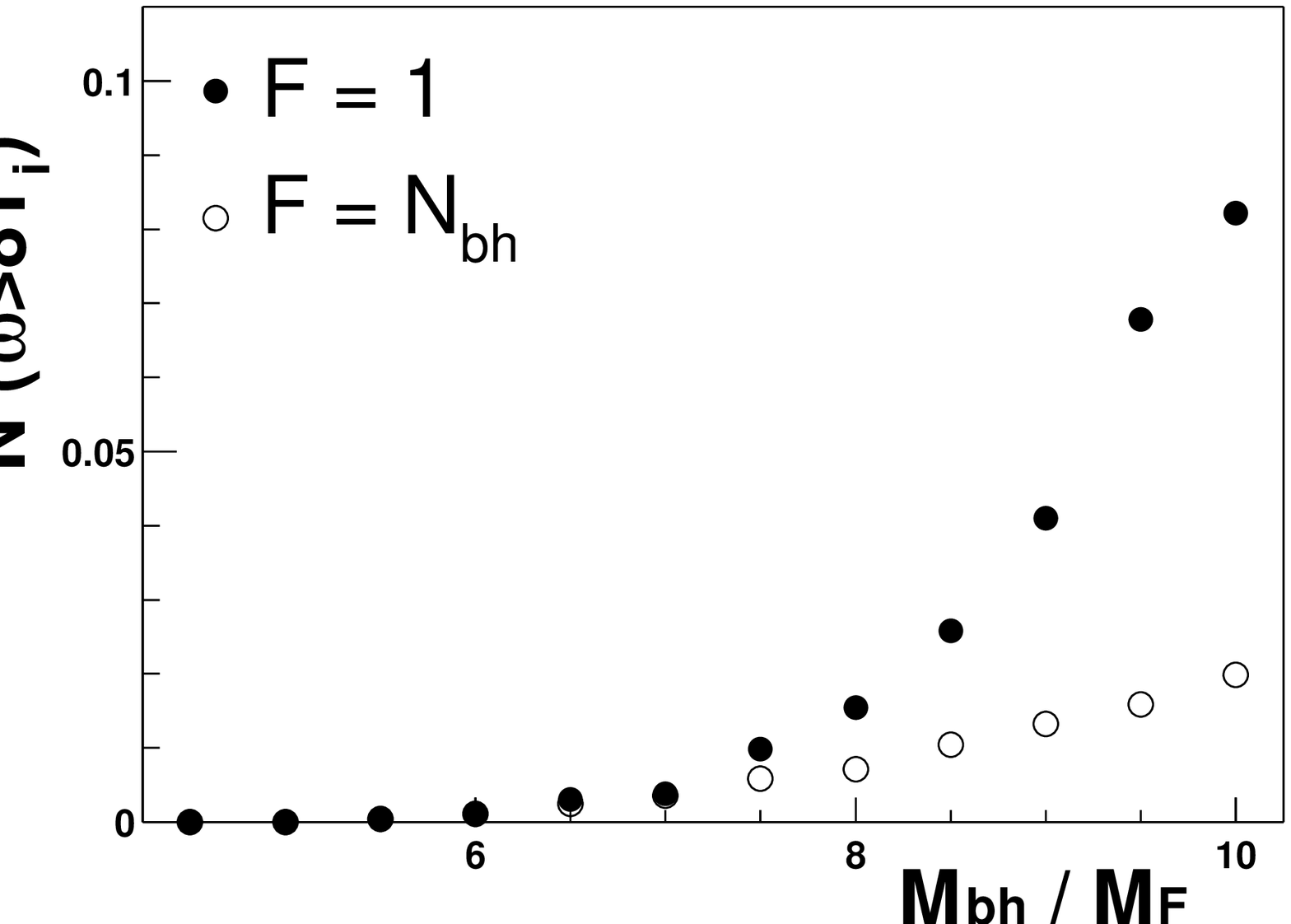}
\epsfxsize 9.0cm
\epsfysize 9.0cm
\epsffile{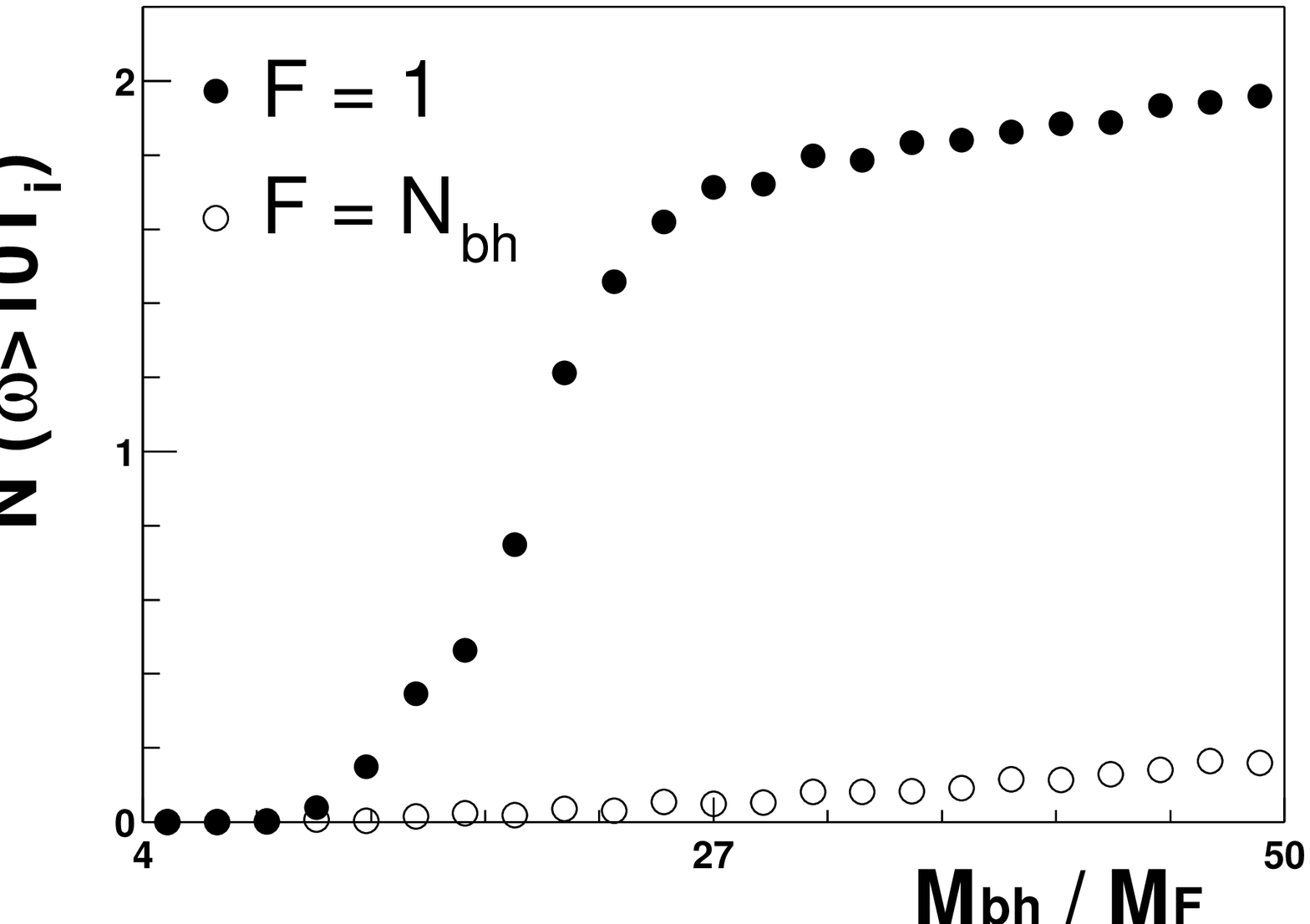}
\vskip 0.15in
\caption{The mean multiplicity $N$ of high
 energy partons per black hole as a function of black hole
mass for the two choices of $\cal F$.  We use 
the thresholds $\omega > 8 T_i$ (top) and $ \omega > 10 T_i$ (bottom), 
where $T_i$ is the initial black hole temperature for $J=0$.
  The distributions are shown for $n=3$.}
\vskip -0.15in
\label{fig:mult}
\end{center}
\end{figure}

{\it Measurement}:  The partons radiated with high energy provide two
 tests of the black hole production dynamics.  
 Figure~\ref{fig:mult} shows the mean multiplicity $N$ of high energy partons
 per black hole
 for ${\cal F}=N_{bh}$ and ${\cal F}=1$ as a function of $M_{bh}$.
  For the mass ranges of 4-10 $M_F$ (the potential 
LHC range) and 5-50 $M_F$ (the potential VLHC range),   we choose 
partons with $\omega > 8 T_i$ and $\omega > 10 T_i$ respectively,
 where $T_i$ is the initial 
temperature for $J=0$.  We find a detectable difference in the multiplicity 
 when $M_{bh} > 8 M_F$, and a substantial difference when
  $M_{bh} > 15 M_F$. 
The sensitivity can be increased beyond the results of this counting experiment
 by analysing the shape of the $\omega$ spectrum at high $\omega$. 

\begin{figure}[hptb]
\begin{center}
\epsfxsize 9.0cm
\epsfysize 9.0cm
\epsffile{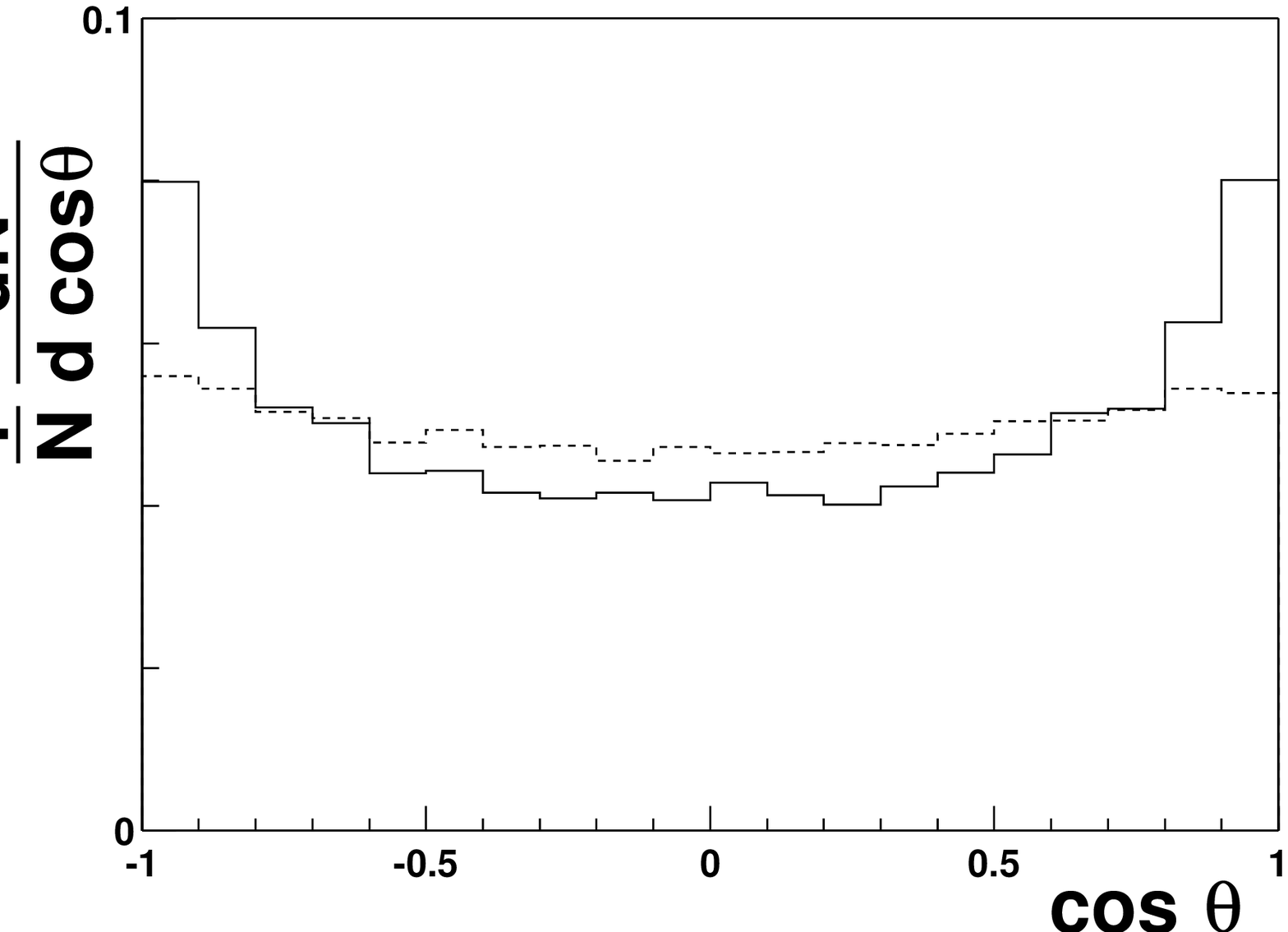}
\vskip 0.15in
\caption{The cos~$\theta$ distribution of the decay partons
 with $\omega > 10 T_i$, for $n$=3 and $M_{bh}$=50$M_F$.  The
 distributions obtained from the two choices of $\cal F$ are 
 compared (solid: ${\cal F} = 1$, dashed: ${\cal F} = N_{bh}$) .}
\vskip -0.15in
\label{fig:costheta}
\end{center}
\end{figure}

The second test of black hole production dynamics is provided by 
the angular 
distribution of the high energy partons.   Figure~\ref{fig:costheta} 
shows the cos~$\theta$~\cite{costheta}
  distributions of the high energy partons
 predicted by the two production models. The  anisotropy when ${\cal F} = 1$
is due to the larger fraction of black holes produced with high $J$ values, 
 resulting
 in Hawking radiation with higher angular momentum.  

Figures~\ref{fig:mult} and~\ref{fig:costheta} show that one can experimentally 
distinguish between the production of high and low angular momentum black 
holes.  We expect this ability to be a powerful test of any theory of quantum 
gravity.  
For example, Voloshin~\cite{voloshin} has used the path-integral approach to 
  introduce in the production cross section 
 the exponential factor $e^{-I_{E} }$, where $I_{E}(M_{bh},J)$
  is the Gibbons-Hawking
 action~\cite{gibbonshawking}. Our results can be used to test this
 prediction and the $J$ dependence of $I_E$. 

 In four dimensions, $I_E = M_{bh} / (2 T_{bh}) $~\cite{gibbonshawking}. If
 we assume that this dependence continues to hold in higher dimensions (up to
 numerical prefactors), we may use the dependence of $T_{bh}$ on $J$ 
 (Eqn.~\ref{temperature}) to extract the $J$ dependence of $I_E$ in
 higher dimensions. For $n>1$, $T_{bh}$ is
 non-zero for any $J$ and has a weak $J$ dependence. 
 For $n > 3$, $T_{bh}$ increases with $J$ for large $J$ so 
 that $I_E$ is smaller 
 at the typical $J \sim R_{bh} M_{bh}$
 than at $J=0$. 
 Therefore in this scenario  the exponential factor $e^{I_E}$ 
 increases the fraction of black holes produced with large $J$ values,
 leading to an enhancement of the multiplicity  and
 anisotropy of high energy Hawking radiation. It would be useful to
 calculate the $4+n$ dimensional Gibbons-Hawking action as a function of
 $M_{bh}$ and $J$, so that these predictions can be tested. 

In conclusion, we have shown that angular momentum $J$ is a salient property
 of collider-produced black holes in space-times with  
extra spatial dimensions. 
In a purely geometrical picture, we have used partial wave
 analysis to calculate the $J$ distribution, showing that 
 the typical value of $J$ is approximately  $R_{bh} M_{bh}$. 
 The large black hole spin is manifest in the increased
 multiplicity
 and anisotropy of the Hawking radiation at high energy. 
 We have shown that black hole phase space considerations can significantly 
 alter the $J$ distribution if the phase space is taken as the number of
 black hole states calculated from the entropy. In a path-integral approach 
 to black hole production, the role of the $4+n$ dimensional Gibbons-Hawking
 action can be tested using the observables we have proposed.

We thank Ronen Plesser, Sailesh Chandrasekharan and Stefan Hofmann 
 for useful discussions. 
We acknowledge support from the U.~S. Department of Energy 
 and the Alfred P. Sloan
 Foundation.


\begin{thebibliography}{99}
\bibitem{planewave}
 The expansion of a plane wave in spherical harmonics $Y_l^m$ is 
$$
e^{i\vec{k} \cdot \vec{r}} =  4\pi \sum_{l m} i^l \; j_l(kr) \; Y_l^m(\hat{r}) \; Y^{m^*}_l(\hat{k}) \quad , \nonumber
$$
where $l(l+1)$ is the eigenvalue of $\hat{L}^2$ ($\hat{L}$ is the two-parton
 relative
 orbital angular momentum operator), $m$ is the eigenvalue of 
$\hat{L}_z$, and $j_l$ are the spherical Bessel functions 
$$ j_l(x) = (-x)^l (\frac{1}{x} \frac{d}{dx})^l (\frac{{\rm sin}~ x}{x}) \quad . $$
\bibitem{finaldecay}
 When the 
 radius shrinks to the minimum allowed value of $R_{min} = M_F^{-1}$, we force
  a final two-body decay that conserves 
energy and angular momentum.
\bibitem{costheta} $\theta$ is the parton decay angle with respect to the
 beam axis in the black hole rest frame.
\end{thebibliography}
\end{document}